\documentclass[prd,
preprintnumbers,amsmath,amssymb]{revtex4}
\usepackage{graphicx}

\DeclareGraphicsExtensions{.pdf,*.jpg}

 \def\be{\begin{equation}}
 \def\ee{\end{equation}}

 \def\r{\rho}
 \def\D{\Delta}
 \newcommand{\la}[1]{\label{#1}}
 \newcommand{\ff}{\frac}
 \newcommand{\vev}[1]{{\left< {#1} \right>}}
 \newcommand{\eq}[1]{(\ref{#1})}
 
   \def\cO{{\cal O}}
 \newcommand{\prt}[1]{{\left( {#1} \right)}}

 \def\s{\sigma}
 
 \def\A{\mathcal{A}}

   \def\R{\mathcal{R}}

 \def\2{\frac{1}{2}}
 \def\4{\frac{1}{4}}
 \def\Tc{{\theta}}

\catcode`\@=11
\def\@citex[#1]#2{%
\if@filesw \immediate \write \@auxout {\string \citation {#2}}\fi
\@tempcntb\m@ne \let\@h@ld\relax \def\@citea{}%
\@cite{%
  \@for \@citeb:=#2\do {%
    \@ifundefined {b@\@citeb}%
      {\@h@ld\@citea\@tempcntb\m@ne{\bf ?}%
      \@warning {Citation `\@citeb ' on page \thepage \space
undefined}}%
      {\@tempcnta\@tempcntb \advance\@tempcnta\@ne%
      \@tempcntb\number\csname b@\@citeb \endcsname \relax%
      \ifnum\@tempcnta=\@tempcntb %Number follows previous--hold on to
it
        \ifx\@h@ld\relax%
          \edef \@h@ld{\@citea\csname b@\@citeb\endcsname}%
        \else%
          \edef\@h@ld{\ifmmode{-}\else--\fi\csname
b@\@citeb\endcsname}%
        \fi%
      \else%   %  non-successor--dump what's held and do this one
        \@h@ld\@citea\csname b@\@citeb \endcsname%
        \let\@h@ld\relax%
      \fi}%
    \def\@citea{,\penalty\@highpenalty\,}%
  }\@h@ld
}{#1}}

\def\@citeb#1#2{{[#1]\if@tempswa , #2\fi}}
\def\@citeu#1#2{{$^{#1}$\if@tempswa , #2\fi }}
\def\@citep#1#2{{#1\if@tempswa , #2\fi}}

\begin{document}
%\preprint{CERN-TH-2019-177}

\title{Entanglement entropy in FRW backgrounds}

\author{D. Giataganas ${}^{a,}$${}^b$}
\email{dimitrios.giataganas@gmail.com}
\author{N. Tetradis ${}^{a}$}
\email{ntetrad@phys.uoa.gr}
\affiliation{ 
${}^{a}$ Department of Physics,
University of Athens,
University Campus,
Zographou 157 84, Greece\\
${}^{b}$ National Sun Yat-sen University, Kaohsiung 80424, Taiwan}
%\date{\today}%

\begin{abstract}
We use holography in order to
study the entanglement entropy for a spherical entangling surface in
a FRW background with an arbitrary time dependence
of the scale factor. The calculation is done in various dimensions, allowing
for nonzero spatial curvature.
The entanglement entropy of a CFT at nonzero temperature in
this background is also considered.
Our approach is based on
coordinate transformations that relate the extremization problem to the one for a
static background, with a careful determination of the UV
cutoff.
We demonstrate the agreement with the expected form of the
entanglement entropy and with various known results in specific cases.
In four dimensions,
apart from the cutoff-dependent terms, we compute and discuss the finite term related to
the expansion rate.

\end{abstract}

%\pacs{}
% PACS, the Physics and Astronomy Classification Scheme.
\maketitle

\section{Introduction}\label{intro}

In recent years there has been an extensive effort to understand strongly
coupled quantum field theories in curved spacetimes. A leading research direction
makes use of the AdS/CFT correspondence \cite{adscft} by
taking advantage of the various slicings of AdS, which result
in different boundary spacetimes. Examples of practical interest include
the de Sitter (dS) spacetime and the Friedmann-Robertson-Walker (FRW) universe.
A  motivation for such considerations comes from cosmology,
where the dynamics of gauge theories takes place in an expanding background.
A strongly coupled process took place
about 10 microseconds after the Big Bang, when the temperature of the universe
crossed the critical temperature of QCD and confinement set in.
It has also been suggested that the dark matter interactions
could require physics beyond the traditional perturbative techniques
\cite{Tulin:2017ara}. For such cases, the holographic description of
field theories in curved backgrounds
can be an important tool.
Another theoretical framework for cosmology within
holography is provided by the dS/CFT correspondence
\cite{Witten:2001kn,Maldacena:2002vr}.

A particularly interesting nonlocal observable is the entanglement entropy, which
has been studied extensively for theories in dS space.
The main reason is that it encodes information on the long-range
correlations of the degrees of freedom and it serves as a probe that can characterize
the state of matter. Moreover, it demonstrates the thermal nature of
field theories in curved spacetimes and approaches the thermal entropy
in certain limits. Direct field theory calculations of the entanglement entropy
are possible mainly in low dimensions
\cite{cardy,discretizer,stefan}
and for non-interacting quantum field theories in higher dimensions \cite{sredincki,casini90,pimentel,Kanno:2014lma,Iizuka:2014rua,Kanno:2016qcc}.
However,
the holographic approach and the Ryu-Takayanagi proposal
\cite{ryu,review}
replace the demanding field theory calculation
of the density matrix by the simpler calculation of extremizing the area
of a surface that starts from the entangling surface and extends into the bulk.
For a boundary with an explicit time dependence, which is our main point of interest,
this surface must be defined
appropriately through a generalization of the proposal for the static
case \cite{rtextension}.
The holographic approach has become an invaluable tool for studies of entanglement
in dS space.
There is an extensive
bibliography with several interesting results, including the connection with
the dS gravitational entropy \cite{strominger,marolfds,pimentel,fischlerds,Chu:2016uwi,Chugiatag,Ghosh:2018qtg,Giataganas:2019wkd,grieninger,correctionsds,vanderSchee,Park,Manu:2020tty,Giataganas:2020lcj}.

In this paper, motivated by the above developments, we extend the holographic
studies of entanglement entropy to general FRW
boundary metrics. We refer to a time-dependent metric with a homogeneous and
isotropic spatial part, in any number of dimensions, as a FRW metric. The time dependence
is encoded in the scale factor, which, with an appropriate choice of the time coordinate,
can be written as a conformal factor multiplying a static metric.
As the boundary metric acts as a source in the
context of the AdS/CFT correspondence, the scale factor is an arbitrary function of
(conformal) time. However, we can consider various forms of this function that correspond
to solutions of the dynamical Einstein equations in known cases, such as the
de Sitter metric in planar coordinates.

The interest in this research direction is not new. Early studies of
entanglement \cite{sredincki} provided the formalism to compute the
entropy via the reduced ground-state density matrix of a free
massless scalar field, by tracing out the degrees of freedom inside an
imaginary sphere and making the analogy with a system of coupled harmonic
oscillators. Based on this technique, explicit calculations for the entanglement
entropy in the FRW universe provided results on its dependence
on the area of the cosmological horizon \cite{Muller:1995mz}.
We improve on these results by considering a spherical entangling surface of arbitrary
radius and computing corrections beyond the leading area term.
In our holographic study we generate the FRW boundary by a set of explicit
coordinate transformations which have as starting point the AdS space in standard
Poincare coordinates.
Such coordinate transformations are well known for a variety of boundary metrics
resulting from appropriate slicings of the bulk AdS space,
see for example \cite{Chu:2016uwi} and \cite{siopsis,nonis}.

Using the appropriate
coordinate transformations, we show that it is possible to map the
extremal surface of the static case
to one in the FRW spacetime. In this way we provide an analytical
way to extremize the area of this surface for the FRW slicing of the AdS space,
a task that seems otherwise intractable
because of the complexity of the problem when approached directly.
Having obtained the analytical form of the extremal surface, we find the UV contributions
to the entanglement entropy in various dimensions and discuss the origin of each term.
Our results justify holographically the known leading dependence of the
entanglement entropy on the area of the entangling surface \cite{Muller:1995mz}.
In 3+1 dimensions, we compute the subleading, logarithmically divergent contribution,
which depends on the number of degrees of freedom (the central charge) of the dual
conformal field theory (CFT), in this case the ${\cal N}=4$ supersymmetric $SU(N)$ gauge theory in the large-$N$
limit. Moreover, we compute finite terms that depend on the expansion rate and
the spatial curvature.

The plan of the paper is as follows: In the following section we
compute the entanglement entropy for a spherical surface in a FRW spacetime with
vanishing spatial curvature. We show how to perform the extremization procedure for
the bulk surface through an appropriate coordinate transformation that reduces the
problem to that for a static boundary. We derive the entanglement entropy in
various dimensions for an arbitrary time dependence of the scale factor.
In section \ref{curvature} we repeat the calculation taking into account
the effect of spatial curvature, by considering a maximally symmetric spatial
part for the boundary metric. In section \ref{temperature} we
compute the entanglement entropy for a (1+1)-dimensional boundary in the
presence of a bulk black hole. The entanglement entropy corresponds to that of
a thermalized CFT in a time-dependent background. We compare with the
entropy in the static case \cite{cardy}.
In section \ref{discussion} we discuss the various terms that appear in the
expression for the entanglement entropy. We show that they agree with the
general expectations, apart from a logarithmically divergent term, whose absence
we explain.
As a side calculation, we compute the equal-time two-point correlation function for
conformal operators in an expanding background with one spatial dimension.
We also discuss the form of the finite term in the entanglement entropy for
realistic cosmologies.

\section{FRW boundary}\label{expansion}

We consider a slicing of $(d+2)$-dimensional
AdS space that results in a boundary
with a $(d+1)$-dimensional Friedmann-Robertson-Walker (FRW) metric.
The bulk metric has the form
\begin{equation}
ds^2_{d+2}
= \frac{1}{z^2} \left[ dz^2
- N^2(\tau,z) d\tau^2
+  A^2(\tau,z) \left( d\rho^2+  \rho^2 \, d\Omega^2_{d-1} \right) \right],
\label{FRW} \end{equation}
where
\begin{eqnarray}
N(\tau,z)&=&a(\tau)\left(1-\frac{-3a'^2(\tau)+2 a(\tau)a''(\tau)}{4a^4(\tau)} \,z^2 \right)
\label{Ntz} \\
A(\tau,z)&=& a(\tau)\left(1-\frac{a'^2(\tau)}{4a^4(\tau)} \,z^2 \right).
\label{Atz}
\end{eqnarray}
All quantities are expressed in terms of the AdS length, which
we set equal to 1.
The boundary metric is spatially flat and its evolution is given
in terms of the conformal time $\tau$. The scale factor $a(\tau)$ is an arbitrary
function of time.
The metric is of the typical Fefferman-Graham form for
an asymptotially AdS space \cite{fg}.
It can also be written using standard Poincare coordinates as
\begin{equation}
ds^2_{d+2}
= \frac{1}{\zeta^2} \left[ d\zeta^2
- dt^2
+  d\rho^2+  \rho^2 \, d\Omega^2_{d-1}  \right].
\label{poincare}
\end{equation}
The metrics (\ref{FRW}) and (\ref{poincare}) are related through the coordinate transformation
\begin{eqnarray}
t(z,\tau)&=&\tau+\frac{2a'(\tau)a(\tau)\,z^2}{-4a^4(\tau)+a'^2(\tau)\,z^2}
\label{tzt} \\
\zeta(z,\tau)&=& \frac{z}{a(\tau)}
\left(1-\frac{a'^2(\tau)}{4a^4(\tau)} \,z^2 \right)^{-1}.
\label{zzt}
\end{eqnarray}
Notice the presence of a singularity at $z=2/H(\tau)$,
where
$H(\tau)=a'(\tau)/a^2(\tau)$ is the expansion rate expressed in terms of
conformal time.
The transformation is well defined for $z<2/H(\tau)$.

We consider a spherical entangling surface $\Sigma$ on the boundary, with
radius $\rho=R$. As $\rho$ is a comoving coordinate, the physical radius of
the surface
grows proportionally to $a(\tau)$, following the general expansion.
The space inside the spherical surface is entangled with the exterior and
we would like to compute the corresponding entanglement entropy at a
given time $T$ using
holography. The explicit time dependence of the configuration
indicates that the relevant holographic framework is provided by the
extension of the Ryu-Takayanagi proposal \cite{ryu,review}
given in ref. \cite{rtextension}.
The entropy is proportional to the area
 \be
 {\rm Area}(\gamma_A))= S^{d-1} I(\epsilon)
 = S^{d-1}\int_\epsilon d\rho\, \rho^{d-1}
 \frac{A^{d-1}(\tau(\rho),z(\rho))}{z^d(\rho)}
 \sqrt{A^{2}(\tau(\rho),z(\rho))
 -N^{2}(\tau(\rho),z(\rho))\left(\frac{d\tau(\rho)}{d\rho} \right)^2
 +\left(\frac{dz(\rho)}{d\rho}\right)^2  },
 \label{areadst} \ee
extremized with respect to the functions $\tau(\rho)$ and $z(\rho)$, with
the boundary conditions $\tau(R,0)=T$ and $z(R,0)=0$.
Here $S^{d-1}$ is the volume of the ($d-1$)-dimensional unit sphere.
The integral diverges near the boundary, so that a cutoff must be imposed {\it
on the bulk coordinate $z$ at $z=\epsilon$.}
The entanglement entropy is given by the relation
\be
S=\frac{{ S^{d-1}}}{4G_{d+2}}I(\epsilon),
\label{entropy} \ee
with $G_{d+2}$ the bulk Newton's constant.
The central charge of the dual CFT is proportional to $1/G_{d+2}$.

Finding the extremum of the area seems a formidable task. However, the
calculation is simplified if one switches to Poincare coordinates, making use
of eqs. (\ref{tzt}), (\ref{zzt}). The functional to be extremized can be written
as
 \be
 {\rm Area}(\gamma_A)= S^{d-1}\int d\rho\, \rho^{d-1}
 \frac{1}{\zeta^{d}(\rho)}
 \sqrt{1
 -\left(\frac{dt(\rho)}{d\rho} \right)^2
 +\left(\frac{d\zeta(\rho)}{d\rho}\right)^2  }.
 \label{areadst} \ee
The solution for the function $t(\rho)$ is trivial: $t(\rho)=T=$constant.
The minimization with respect to $\zeta(\rho)$ is the standard one for
a spherical entangling surface in the context of the Ryu-Takayanagi
proposal for a static boundary.
The minimal surface for an entangling surface with constant comoving radius $R$
is determined by the relation $\zeta(\rho)=\sqrt{R^2-\rho^2}$.
Combining these facts, the extremal surface corresponding to an entangling surface of
comoving radius $R$ at a time $T$ on the boundary is given by the implicit relations
\begin{eqnarray}
T&=&\tau+\frac{2a'(\tau)a(\tau)\,z^2}{-4a^4(\tau)+a'^2(\tau)\,z^2}
\label{soltau} \\
\sqrt{R^2-\rho^2}&=& \frac{z}{a(\tau)}
\left(1-\frac{a'^2(\tau)}{4a^4(\tau)} \,z^2 \right)^{-1}
\label{solzeta}
\end{eqnarray}
for the functions $\tau(\rho)$ and $z(\rho)$.

The integration for the area of the extremal surface can be performed by using $\zeta$
as an independent variable.
The area becomes \cite{ryu,review}
\be
{\rm Area}(\gamma_A)=
 S^{d-1}I(\epsilon) =
 S^{d-1}\int_{\epsilon_\zeta(T)/R}^1 dy \frac{(1-y^2)^{(d-2)/2}}{y^d},
\label{areaeps} \ee
where $y=\zeta/R$.
The area must be regulated by imposing a cutoff on $\zeta$
near the boundary. This cutoff results from the cutoff $\epsilon$ imposed on the
Fefferman-Graham coordinate $z$.
For $\epsilon\to 0$, we obtain from eq. (\ref{soltau})
\be
\tau(\epsilon)=T+\frac{a'(T)}{2a^3(T)}\epsilon^2
\label{tauex} \ee
near the boundary.
Using this expression in order to expand $a(\tau)$ in eq. (\ref{zzt})
we find that the cutoff on $\zeta$ can be written as
\be
\epsilon_\zeta(T)=\frac{\epsilon}{a(T)} \left(1-\frac{1}{4}H^2(T)\epsilon^2 \right),
\label{cutoff} \ee
where we have defined the Hubble parameter
$H(T)=a'(T)/a^2(T)$ in terms of the conformal time on the boundary.

It is remarkable that the whole dependence of the entropy on $T$
enters through $\epsilon_\zeta(T)$. This cutoff results from the fundamental
cutoff $\epsilon$ that regulates the UV divergences of the theory.
%The singular terms in the entanglement entropy reflect the UV physics
%and are independent of the expansion of the background.
%In this sense, the regularization must be
%performed {\it with the same value of $\epsilon$} for any form of
%the scale factor $a(T)$.
For $\epsilon \to 0$
the integral of (\ref{areaeps}) can be evaluated in various dimensions.
Following the procedure outlined above, we find
\begin{eqnarray}
d=3&~~~~~~~~~~~~~&
I(\epsilon)=\frac{a^2(T)R^2}{2\epsilon^2}+\frac{1}{2}\log \left(\frac{ \epsilon}{2a(T)R} \right)
+\frac{1}{4} a^2(T)R^2H^2(T)-\frac{1}{4}
+{\cal O}(\epsilon^2)
\label{expandI3frw} \\
d=2&~~~~~~~~~~~~~&
I(\epsilon)=\frac{a(T)R}{\epsilon}-1
+{\cal O}(\epsilon^1)
\label{expandI2frw} \\
d=1&~~~~~~~~~~~~~&
I(\epsilon)=-\log \left(\frac{ \epsilon}{2a(T)R} \right)
+{\cal O}(\epsilon^2).
\label{expandI1frw}
\end{eqnarray}
The time dependence of the scale factor $a(T)$ is arbitrary,
as the boundary metric does not result from dynamical equations of motion.
However, the above expressions are applicable to physical FRW cosmologies
with metrics that have a dynamical origin.
In $(3+1)$-dimensional
cosmology, with the expansion driven by matter with an
equation of state $p=w \varepsilon$, we would have $a(T)\sim T^{2/(1+3w)}$ for $w>-1/3$.
For $w<-1/3$, the conformal time is negative. In the
particular case of a cosmological constant, with $w=-1$ and constant $H$, we
have $a(T)=-(H T)^{-1}$, where $T$ takes values between $-\infty$ and 0.

\section{Spatial curvature} \label{curvature}

The analysis can be generalized for a boundary FRW metric with spatial
curvature.
The bulk metric has the form
\begin{equation}
ds^2_{d+2}
= \frac{1}{z^2} \left[ dz^2
- N^2(\tau,z) d\tau^2
+  A^2(\tau,z) \left( \frac{d\rho^2}{1-\frac{k}{R_0^2}\,\rho^2}+  \rho^2 \, d\Omega^2_{d-1} \right) \right],
\label{FRWc} \end{equation}
with
\begin{eqnarray}
N(\tau,z)&=&a(\tau)\left(1-\frac{1}{4}\left(
{-3\frac{a'^2(\tau)}{a^4(\tau)}-\frac{k}{a^2(\tau)R^2_0}
+2 \frac{a''(\tau)}{a^3(\tau)}}\right) \,z^2 \right)
\label{Ntzc} \\
A(\tau,z)&=& a(\tau)\left(1-\frac{1}{4}\left(\frac{a'^2(\tau)}{a^4(\tau)}+\frac{k}{a^2(\tau)R^2_0} \right) z^2 \right),
\label{Atzc}
\end{eqnarray}
with $k=0,\pm 1$, depending on the spatial curvature of the boundary. The parameter
$R_0$ sets the scale of the spatial curvature.
This metric can also be rewritten as
\begin{equation}
ds^2_{d+2}
= \frac{1}{\zeta^2} \left[ d\zeta^2
- \left(1+\frac{1}{4}\frac{k}{R^2_0}\zeta^2 \right)^2 dt^2
+ \left(1-\frac{1}{4}\frac{k}{R^2_0}\zeta^2 \right)^2 \left( \frac{d\rho^2}{1-\frac{k}{R^2_0} \rho^2}
+  \rho^2 \, d\Omega^2_{d-1} \right)  \right].
\label{poincarec}
\end{equation}
The two metrics are related through a coordinate transformation that does not involve $\rho$ and the
angular variables \cite{siopsis}. The relation between the bulk coordinates $z$ and $\zeta$
is given by the expression
\be
\zeta(z,\tau)\left( 1-\frac{1}{4}\frac{k}{R^2_0}\zeta^2(z,\tau) \right)^{-1} = \frac{z}{a(\tau)}
\left(1-\frac{1}{4}\left(\frac{a'^2(\tau)}{a^4(\tau)}+\frac{k}{a^2(\tau)R^2_0} \right)z^2\right)^{-1}.
\label{zztc}
\ee
It is clear from this expression that the curvature effects
disappear for large $R_0$, as expected. For $R_0 \gg  1/(a(T)H(T))$
we recover eq. (\ref{zzt}).
In order to derive the relation between the time coordinates $\tau$ and $t$,
one must consider the cases $k=0,\pm 1$ separately. The
case $k=0$ was considered earlier.
For the other two cases, we find
\begin{eqnarray}
t(z,\tau)&=\tau+R_0 \tan^{-1}\left(\frac{2 a'(\tau)a(\tau)\,R_0\,z^2}{-4 a^4(\tau)\,R_0^2
+\left( -a^2(\tau)+ a'^2(\tau)\,R^2_0 \right)z^2} \right)
~~~~~~&{\rm for~}k=1
\label{tztt1} \\
t(z,\tau)&=\tau+R_0 \tanh^{-1}\left(\frac{2 a'(\tau)a(\tau)\,R_0\,z^2}{-4 a^4(\tau)\,R_0^2
+\left( a^2(\tau)+ a'^2(\tau)\,R^2_0 \right)z^2} \right)
~~~~~~&{\rm for~}k=-1.
\label{tztt2}
\end{eqnarray}

The extremal surface is
most easily determined in the system of coordinates of eq. (\ref{poincarec}).
Following our earlier logic, it becomes apparent that the extremal surface is $t$-independent and
corresponds to the minimization of the functional
\be
{\rm Area}(\gamma_A)= S^{d-1}\int_{\epsilon_\zeta(T,R)} d\rho\, \rho^{d-1}
\frac{\left(1-\frac{1}{4}\frac{k}{R^2_0} \zeta^2\right)^{d-1}}{\zeta^d}
\sqrt{\frac{\left(1-\frac{k}{4}\frac{1}{R^2_0}\zeta^2\right)^{2}}{1-\frac{k}{R^2_0} \rho^2}
+\left(\frac{d\zeta(\rho)}{d\rho}\right)^2  }.
\label{areadsc} \ee
The solutions of the minimization problem can be summarized as
\be
\zeta(\rho)=2R_0 \sqrt{\frac{-k+k\sqrt{\frac{R^2_0-k\rho^2}{R^2_0-kR^2}}}{
1+\sqrt{\frac{R^2_0-k\rho^2}{R^2_0-kR^2}}}}
~~~~~~{\rm for~}k=\pm 1.
\label{solwc}
\ee
For $R/R_0 \ll 1$, the solution becomes
$\zeta=\sqrt{R^2-\rho^2}$, and
the maximal value of the parameter $\zeta$ is equal to $R$.
%corresponding to the known expression for
%$k=0$ \cite{ryu,review}.
Near the boundary, eq. (\ref{zztc}) indicates that to leading order
$\zeta(z,\tau)=z/a(\tau)$.
Therefore, the combination $\zeta^2/R^2_0$ in the lhs of this equation has
a maximal value equal to $R^2/R_0^2$, similarly to the combination
$z^2/(a^2R_0^2)$ in the rhs. On the other hand, the combination
$a'^2z^2/a^4$ in the rhs has a maximal value equal to $H^2a^2R^2$, and can be
substantial even for $R/R_0\ll 1$. We reach the expected
conclusion that the curvature effects become negligible at
short length scales, even though the expansion of the background
remains relevant.

The cutoff $\epsilon_\zeta$ can be related to the cutoff imposed on the
Fefferman-Graham coordinate $z=\epsilon$ by expanding eqs. (\ref{zztc}), (\ref{tztt1}), (\ref{tztt2}) up to order $\epsilon^3$
and setting $t=T$, with $T$ the value that the function $\tau(\rho)$ on the extremal surface takes when the boundary is approached.
This returns eq. (\ref{cutoff}).
The integral of eq. (\ref{areadsc}) can be evaluated in various dimensions.
We find
\begin{eqnarray}
d=3&~~~~~~~~~~~~~&
I(\epsilon)=\frac{a^2(T)R^2}{2\epsilon^2}+\frac{1}{2}\log \left(\frac{ \epsilon}{2a(T)R} \right)
+\frac{1}{4}a^2(T)R^2H^2(T) +\frac{1}{4}\left( k\frac{R^2}{R_0^2}-1\right)
+{\cal O}(\epsilon^2)
\label{expandI3frwk} \\
d=2&~~~~~~~~~~~~~&
I(\epsilon)=\frac{a(T)R}{\epsilon}-1
+{\cal O}(\epsilon^1)
\label{expandI2frwk} \\
d=1&~~~~~~~~~~~~~&
I(\epsilon)=-\log \left(\frac{ \epsilon}{2a(T)R} \right)
+{\cal O}(\epsilon^2).
\label{expandI1frwk}
\end{eqnarray}
The comparison of eqs. (\ref{expandI3frw}) and (\ref{expandI3frwk})
reveals the influence of the spatial curvature on the entanglement entropy.
The curvature effects are confined to the last finite term.
The absence of a logarithmic term depending on the curvature will be discussed
in the last section.

\section{Temperature}\label{temperature}

In order to include an energy scale in the context of an expanding background
one must abandon the simple framework of a pure AdS bulk. The simplest
generalization involves the presence of a bulk black hole of mass $\mu$.
The dual CFT is thermalized at a temperature $\Tc_0=\sqrt{\mu}/(2\pi)$.
We consider the simplest example of a (2+1)-dimensional bulk, for which
the analytical expressions are transparent and provide an intuitive picture
of the effect of the expansion.
The higher-dimensional cases require a numerical treatment which we defer to future work.

The non-rotating BTZ black hole \cite{BTZ} is a solution of the Einstein equations in 2+1 dimensions in the presence of a
negative cosmological constant:
$\Lambda_3 = -1$.
The metric can be written in Schwarzschild coordinates as
\begin{equation}
\label{eqmetric}
ds^2 = -f(r) dt^2 + \frac{dr^2}{f(r)} + r^2 d\phi^2,\ \ \ \ \ \ f(r) = r^2- \mu.
\end{equation}
The Hawking temperature of the black hole is
$\Tc_0 = \sqrt{\mu}/(2\pi)$. In order for the above solution
to have the properties of a black hole, the coordinate
$\phi$ must be periodic, with period equal to $2\pi$. However, the presence of
a bulk horizon even for  noncompact $\phi$ implies that the interpretation of
the dual theory as a thermalized CFT also holds for the more general case.

The above metric can also be expressed as \cite{siopsis,nonis}
\begin{equation}
ds^2
= \frac{1}{z^2} \left[ dz^2
- N^2(\tau,z) d\tau^2
+  A^2(\tau,z)d\phi^2 \right],
\label{FRWm} \end{equation}
where
\begin{eqnarray}
N(\tau,z)&=&a(\tau)\left(1-\frac{\mu\, a(\tau)^2-3a'^2(\tau)+2 a(\tau)a''(\tau)}{4a^4(\tau)} \,z^2 \right)
\label{Ntzm} \\
A(\tau,z)&=& a(\tau)\left(1+\frac{\mu\, a(\tau)^2 -a'^2(\tau)}{4a^4(\tau)} \,z^2 \right).
\label{Atzm}
\end{eqnarray}
The metrics (\ref{eqmetric}) and (\ref{FRWm}) are related through the coordinate transformation
\begin{eqnarray}
t(z,\tau)&=&\tau+
\frac{1}{2\sqrt{\mu}}\log\left[
\frac{4 a^4-\left(\sqrt{\mu}\, a(\tau)+a'(\tau) \right)^2 z^2 }{
4 a^4-\left(\sqrt{\mu}\, a(\tau)-{a'(\tau)}\right)^2 z^2 }
\right]
\label{tztm} \\
r(z,\tau)&=& \frac{a(\tau)}{z}
\left(1+\frac{\mu\, a(\tau)^2- a'^2(\tau)}{4a^4(\tau)} \,z^2 \right).
\label{zztm}
\end{eqnarray}

For the metric (\ref{FRWm}),
it is instructive to consider the
stress-energy tensor of the dual CFT on the time-dependent boundary, as
determined via holographic renormalization \cite{skenderis}.
We obtain the energy density and pressure
\begin{eqnarray}
\label{eq3te}
\rho&=&-\langle T_{~t}^{t} \rangle=
\frac{1}{16\pi G_3}\left( \frac{\mu}{a^2} - \frac{{a'}^2}{a^4}\right)
\\
P& =& \langle T_{~x}^{x} \rangle =
\frac{1}{16\pi G_3} \left( \frac{\mu}{a^2}
+ \frac{-3{a'}^2+2a{a''}}{a^4} \right).
\label{eq3tp}
\end{eqnarray}
The terms proportional to $\mu/a^2$ can be interpreted as the
thermal energy density
and pressure of a CFT at a temperature $\Tc(T)=\Tc_0/a(T)$.

We turn next to the entanglement entropy of the CFT in the time-dependent
setting of eq. (\ref{FRWm}).
We consider a segment of
length $2R$ for the comoving coordinate $\phi$
on the boundary. Similarly to the previous sections,
the minimization of the area functional can be performed by
switching to the coordinates $(t,r)$, for which the minimal curve
has the trivial time dependence $t(\phi)=T$=constant. It is obvious from
eq. (\ref{tztm}) that $T$ corresponds to the value that
the function $\tau(\phi)$ on the minimal surface takes when the boundary
is reached.
The calculation of the minimal area and the resulting entanglement
entropy for the static case has been
performed in ref. \cite{rtextension}. It reproduces the well known
result \cite{cardy}
\be
S=\frac{1}{2G_3}\log\left(\frac{1}{\pi \,\Tc_0\, \epsilon_r}
\sinh(2\pi\, \Tc_0 \, R) \right).
\label{thermalentr} \ee
The cutoff $\epsilon_r$ has been imposed on the bulk coordinate
at $r=1/\epsilon_r$.
This cutoff can be related to the cutoff $\epsilon$ that we must
impose on the Fefferman-Graham coordinate $z$ for the metric (\ref{FRWm}).
We first observe that eq. (\ref{tztm}) with $t=T$ implies that
$\tau=T+{\cal O}(\epsilon^2)$ at $z=\epsilon$ near the boundary.
We then obtain from eq. (\ref{zztm}) that
$\epsilon_r=\epsilon/a(T)+{\cal O}(\epsilon^3)$.
The higher-order corrections in this expression can be neglected
for a $(1+1)$-dimensional
boundary because the divergences in the area of the minimal surface are
not strong. However, this is not the case in higher dimensions,
in which the corrections of order $\epsilon^2$ in the expansion of the function
$\tau(\phi)$ on the extremal surface must be preserved,
as was done in eq. (\ref{tauex}). Expressing eq. (\ref{thermalentr}) in terms
of $\epsilon$ we obtain
\be
S=\frac{1}{2G_3}\log\left(\frac{a(T)}{\pi \,\Tc_0\, \epsilon}
\sinh(2\pi\,\Tc_0\, R) \right)
=\frac{1}{2G_3}\log\left(\frac{1}{\pi\, \Tc(T)\, \epsilon}
\sinh(2\pi\, \Tc(T)\,  a(T)R) \right)
\label{thermalentropy} \ee
for the time-dependent background.
For vanishing temperature $\Tc_0\to 0$ we recover eq. (\ref{expandI1frw}).
The second equality above allows us to deduce the variation of the
entanglement entropy for a fixed physical length $2a(T)R$ of the
segment. Of course, fixing the physical length
implies that the comoving length $2R$ shrinks as $1/a(T)$
for increasing $a(T)$.
For a fixed physical length and a temperature that is reduced by the
expansion as $1/a(T)$, the entanglement entropy drops and asymptotically
reaches a constant value given again by eq. (\ref{expandI1frw}), but
now for constant $a R$.
%The opposite
%happens during a contraction, corresponding to decreasing $a(T)$.

\section{Discussion and conclusions}\label{discussion}

We would like to confirm our results by comparing them with
the general expectations for the form of the entanglement entropy, as well
as with known results in specific cases.
For $d=3$, as has been observed in ref. \cite{pimentel},
we expect that the entropy for a spherical entangling surface in a time-dependent background with a maximally symmetric spatial part, has the form
\be
S=C_1\frac{\A}{\epsilon^2}+\left(C_2+C_3 \A \right)\log(\epsilon)+C_4\log(\A)+C_5\A.
\label{generalpar}\ee
The proper area of the spherical surface is $\A(T,R)= S^2 a^2(T) R^2=4\pi  a^2(T) R^2$ and
the expansion rate $H(T)=a'(T)/a^2(T)$. The terms involving $C_1$, $C_2$, $C_4$
would be present also for a static flat background. The terms involving $C_3$, $C_5$
have an explicit dependence on the expansion rate and the spatial curvature.
We have neglected finite terms that are mere constants.
In order to check the consistency with the above expression, we rewrite
eqs. (\ref{entropy}), (\ref{expandI3frwk}) as
\be
S=\frac{1}{8 G_5 \epsilon^2}\A+\frac{\pi }{2G_5}\log \left(\frac{ \epsilon}{\sqrt{\A}} \right)
+\frac{1}{16G_5} \left(H^2+\frac{k}{a^2 R_0^2} \right) \A,
\label{entropycompare} \ee
where again we have neglected various constants.
The first term is the standard area term, which can be put in a form reminiscent of
the Bekenstein-Hawking entropy \cite{bekhawk}
by defining an effective Newton's constant
$G^{\rm eff}_4=2G_5 \epsilon^2$ \cite{strominger,correctionsds}.
The coefficient of the second term involves the combination that is proportional to
the central charge of the dual CFT, in this case
the ${\cal N}=4$ supersymmetric $SU(N)$ gauge theory in the large-$N$ limit: $\pi/(2G_5)=N^2$.
For this particular CFT, the coefficients $C_2$ and $C_4$ are related. In particular,
we have $C_2=-2C_4=N^2$. The last term in eq. (\ref{entropycompare}) determines
the coefficient $C_5$ in eq. (\ref{generalpar}), displaying its explicit dependence
on the expansion rate and the spatial curvature.
Notice also that the form of the logarithmic term is consistent with
the direct calculation of the entanglement entropy in dS space
for a theory of a massless,
minimally coupled scalar field, for which the coefficient is equal to 1/90
\cite{pimentel}.

The absence of the logarithmic term proportional to $C_3$ in eq. (\ref{entropycompare})
is conspicuous. The reason can be traced to the special properties of the dual CFT
and the spherical symmetry of the entangling surface. It is sufficient to
understand the absence of this term for a static curved space on
the boundary, whose metric is given by eq. (\ref{poincarec}) for $\zeta=0$.
The time-dependent case, corresponding to the boundary of the metric (\ref{FRWc}),
is conformally related to the static one, and displays the same feature.
It was shown in ref. \cite{solodukhinc3} that, in 3+1 dimensions,
the entanglement entropy of
the ${\cal N}=4$ supersymmetric $SU(N)$ gauge theory in a curved background
includes the term
\be
C=-\frac{N^2}{24\pi}\int_\Sigma \left(3\R_{aa}-2\R-\frac{3}{2} k_a k_a \right)\log(\epsilon),
\label{C3term} \ee
where the integral is taken over the entangling surface $\Sigma$ of radius $R$.
Here $\R$ is the curvature scalar, $\R_{aa}=\R_{\mu\nu}n^\mu_an^\nu_a$ the
projection of the Ricci tensor on the space orthogonal to $\Sigma$, $n^\mu_a$, $a=1,2$
two unit vectors on this space, and $k_a$ the trace of the extrinsic curvature.
Summation over $a$ is assumed, with the appropriate sign for a timelike vector.
For the case at hand, we can take $n_1=(1,0,0,0)$, $n_2=(0,\sqrt{1-kR^2/R^2_0},0,0)$.
The first vector does not contribute to (\ref{C3term}), and we have
$\R=6k/R_0^2$, $\R_{aa}=2k/R_0^2$ and $k_ak_a=4(1/R^2-k/R^2_0)$.
Substitution in eq. (\ref{C3term}) gives
\be
C=\frac{N^2}{24\pi}\int_\Sigma \frac{6}{R^2}\log(\epsilon)=N^2\log(\epsilon)
=C_2\log(\epsilon),
\label{CC3} \ee
showing that $C_3=0$.

For $d=1$ the result of eq. (\ref{expandI1frwk}) was derived
in ref. \cite{Chugiatag} for the case of
dS space using holography. An interesting
first-principles calculation of the
entanglement entropy for a time-dependent situation is presented in ref. \cite{stefan}.
It concerns the expanding
light-cone geometry for the massless Schwinger model of quantum electrodynamics in
1+1 spacetime dimensions. It is found that the entanglement entropy in a finite
rapidity interval $\Delta\eta$ is equivalent
to that of a (1+1)-dimensional conformal field theory at a finite temperature $T$
that scales as $ 1/\tau$ in terms of proper time. The entanglement entropy
computed in
%given by eq. (4.37) of
ref. \cite{stefan} has the
exact form of eq. (\ref{thermalentropy}), with $R=\Delta\eta/2$, $a(\tau)= \tau$
and $\theta_0=1/(2\pi)$.

The result (\ref{expandI1frwk}) for the entanglement entropy for $d=1$
is related to
the form of the two-point correlation function of conformal operators in a
FRW background.
A scalar field of mass $m$ in the bulk of $AdS_3$ corresponds to
an operator of the dual theory with conformal dimension $\D=1+\sqrt{1+m^2}$.
The bulk propagator between points that approach the boundary
is approximated in the semiclassical limit by
\be\la{cor}
G(x,x')=e^{-\D L_g},
\ee
where $L_g$ is the length of the bulk geodesic joining the two boundary points
at $\rho=-R$ and $\r=R$.
The expression \eq{cor} corresponds to the dual correlator in the
large N limit \cite{Balasubramanian:1999zv,Banks:1998dd}. The length of the bulk
geodesic also determines the entanglement entropy and leads to
(twice) the result of eq. (\ref{expandI1frwk}).
This implies that
the two-point function of the conformal operator has the dependence
\be
\vev{\cO\left( -R,T \right) \cO\left(R,T\right)}\sim \prt{\ff{1}{2a(T) R}}^{2\D},
\ee
in agreement with the result for the equal-time
correlator obtained in ref. \cite{Koyama:2001rf}. This is expected,
as the correlator is fully determined by the conformal invariance and
should be proportional to $\s^{-2\D}$, where $\s$ is the FRW invariant distance.
However, the agreement serves as a crosscheck of the validity of our calculation.

Our result for the entanglement entropy in a (3+1)-dimensional background,
given by eqs. (\ref{entropy}), (\ref{expandI3frwk}) or eq. (\ref{entropycompare}),
 displays several interesting features:
\begin{enumerate}
\item
The divergent terms have the same form as for a static background, with the
radius of the entangling surface corresponding to the physical radius
$a(T)R$ that determines the proper area of the surface.
This indicates that the divergences are associated with
the entanglement of UV degrees of freedom very close to the entangling surface, for
which the proper area is the only relevant parameter.
A possible logarithmic divergence that would depend on the curvature of the
background related to the expansion rate
is absent for the particular theory that we considered, as we
explained above.
\item
The spatial curvature gives a finite contribution
that depends quadratically on
the physical radius, so that the effect is proportional to the proper area.
This contribution vanishes when the ratio of the physical radius to the curvature radius
of the background goes to zero.
\item
The finite term involving the expansion rate in eq. (\ref{expandI3frwk})
accounts for contributions to the entanglement entropy from regions not confined in
the vicinity of the entanglement surface. It displays interesting features
for choices of the scale factor $a(T)$ that correspond to realistic FRW cosmologies:
\begin{itemize}
\item
The dependence of the finite term
on the square of the expansion rate means that it has the same value
both for expanding and contracting cosmologies.
\item
As we discussed at the end of section \ref{expansion}, the choice
$a(T)=-(HT)^{-1}$, with constant $H$ and negative conformal time $T$,
corresponds to a de Sitter background. In this case,
the finite term is still proportional to the area of the entangling surface.
The absence of an effect proportional to the volume enclosed by the entangling
surface, which would scale $\sim a^3(T)$ for a fixed comoving radius $R$, could
be attributed to the rapid expansion that hinders the entanglement.
Notice that, for constant $H$, the transformations of eqs. (\ref{tzt}), (\ref{zzt}) are
well defined for $z<2/H$. For a minimal surface described by
eqs. (\ref{soltau}), (\ref{solzeta}) with fixed values of $R$ and $T$, the maximal
value $z_m$ of $z(\rho)$ on the surface is obtained for $\rho=0$.
Increasing $R$ for fixed $T$ makes
$z_m$ increase, up to the value $z_m=2/H$ which is attained when $R=-T$.
This last relation can be rewritten as $a(T)R=1/H$, which indicates that the
entangling surface extends up to the physical horizon radius. The procedure that
we have followed allows the study of subhorizon entangling surfaces, for which
the finite term in the entanglement entropy satisfies
$a^2(T)R^2H^2\leq 1$.
\item
For a cosmological expansion driven by matter with an equation of state
$p=w\varepsilon$ with $w>-1/3$, the scale factor evolves as $a(T)\sim T^{2/(1+3w)}$.
At very early times $T\to 0^+$, the finite contribution to the
entanglement entropy within a comoving radius $R$ scales as
$a^2(T)R^2H^2(T)\sim T^{-2}$.
%This means that the finite contribution to
%the entanglement entropy grows in this limit and falls off at later times.
Its growth at early times could be attributed to
physical distances between adjacent points being small in this limit, so
that the entanglement becomes stronger.
%Of course, this behavior is limited to length scales and the corresponding times
%above the cutoff $\epsilon$.

\end{itemize}
\end{enumerate}

As a final remark, we emphasize the general applicability of our results for
any form of the scale function $a(T)$ and any number of dimensions. It is
also noteworthy that the total dependence of the entanglement entropy
on the expansion rate arises through the effective cutoff
of eq. (\ref{cutoff}). The reason for this feature is that the problem
of finding the extremal surface can be reduced to the one in the
static case through an appropriate coordinate transformation.
In principle, the area of an extremal surface should
be independent of the system of coordinates.
However, the presence of a divergence requires an additional assumption about
its regularization. We assumed that the relevant cutoff is the one
imposed on the bulk coordinate $z$ in a Fefferman-Graham expansion that starts
with the time-dependent metric of interest on the boundary.
The coordinate transformation determines uniquely the cutoff in the
corresponding static case, which now encodes all the information about the
original background. It is remarkable that this simple scheme
reproduces the entanglement entropy in all the known special cases, and
also provides information about the finite terms in the entanglement entropy.
Finding a transformation to the static problem may be more
difficult when the assumption of spatial homogeneity and isotropy is
relaxed. However, it seems to be the most efficient way for dealing with the
general case, without assuming a specific form of the scale factor.

\section*{Acknowledgments}
We would like to thank S. Floerchinger, E. Kiritsis and W. van der Schee for
useful discussions.
The research of N. Tetradis was supported by the Hellenic Foundation for
Research and Innovation (H.F.R.I.) under the “First Call for H.F.R.I.
Research Projects to support Faculty members and Researchers and
the procurement of high-cost research equipment grant” (Project
Number: 824).

%\newpage

\end{document}